\documentclass{aa}

\usepackage[utf8]{inputenc}
\usepackage{graphicx}
\usepackage{amsmath, amssymb}
\usepackage{booktabs}
\usepackage{hyperref}
\usepackage{enumitem}
\usepackage{dblfloatfix}

\renewenvironment{thebibliography}[1]{%
  \begin{oldthebibliography}{#1}%
  \setlength{\itemsep}{\baselineskip}%
}{\end{oldthebibliography}}

\title{A photometric classifier for tidal disruption events in \textit{Rubin} LSST}

\author{
    Kunal Bhardwaj\thanks{Email: bhardwaj@fzu.cz}
    \and Asen Christov
    \and Sergey Karpov
}

\institute{
    Institute of Physics of the Czech Academy of Sciences, Praha, Czech Republic
}

\titlerunning{Photometric classifier for TDEs in LSST}
\authorrunning{Bhardwaj et al.}

\date{Received ZZZ; revised YYY; accepted XXX}

\begin{document}

\abstract
{Tidal disruption events (TDEs) are astrophysical phenomena that occur when stars are disrupted by supermassive black holes. The \textit{Vera C. Rubin} Observatory Legacy Survey of Space and Time (LSST), with its unprecedented depth and cadence, will detect thousands of TDEs, creating the need for robust photometric classifiers capable of efficiently distinguishing these events from other extragalactic transients.}
{We developed and validated a machine learning pipeline for photometric TDE identification in LSST-scale datasets. Our classifier is designed to provide high precision and recall, enabling the construction of reliable TDE catalogs for multi-messenger follow-up and statistical studies.}
{Using the second Extended LSST Astronomical Time Series Classification Challenge (ELAsTiCC2) dataset, we fit Gaussian processes (GPs) to light curves for feature extraction (e.g., color, rise and fade times, and GP length scales). We then trained and tuned boosted decision-tree models (XGBoost) with a custom scoring function that emphasizes the high-precision recovery of TDEs. Our pipeline was tested on diverse simulations of transient and variable events, including supernovae, active galactic nuclei, and superluminous supernovae.}
{We achieve high precision (up to 95\%) while maintaining competitive recall (about 72\%) for TDEs, with minimal contamination from non-TDE classes. Key predictive features include post-peak colors and GP hyperparameters that reflect the characteristic timescales and spectral behaviors of TDEs.} 
{Our photometric classifier provides a practical and scalable approach to identifying TDEs in forthcoming LSST data. By capturing essential color and temporal properties through GP-based feature extraction, it enables the efficient construction of clean TDE candidate samples. Future refinements will incorporate real data and additional features (e.g., photometric redshifts), further enhancing the reliability and scientific impact of this classification framework.}


\maketitle\textit{a}

\section{Introduction}
\label{sec:intro}

Tidal disruption events (TDEs) occur when stars venture close to supermassive black holes (SMBHs) in the center of galaxies, resulting in the disintegration of stars as their self-gravity is overwhelmed by the tidal forces of SMBHs. Consequently, the disrupted stellar debris is flung around the SMBH, and some fraction of it, depending on the type of disruption, remains bound and eventually circularizes to form an accretion disk; the rest escapes the gravitational field of the SMBH \citep{rees_tidal_1988, evans_tidal_1989}. The emanating radiation spans the entire electromagnetic spectrum, although there are still ongoing discussions about the exact physical mechanisms involved (see \citealt{gezari_tidal_2021} for a recent review of TDEs). These cataclysmic events are not only scientifically rich in their own right but also serve as potential accelerators of ultra-high-energy cosmic rays \citep[][]{piran_ultra_2023, biehl_tidally_2018, farrar_tidal_2014} and sources of high-energy neutrinos \citep{stein_tidal_2021, yuan_at2021lwx_2024, vanvelzen_establishing_2024}. Currently, only around 100 to 150 TDEs have been discovered, primarily in optical surveys, and expanding that sample will be crucial for revealing additional multi-messenger correlations and potentially shedding light on the origins of some ultra-high-energy cosmic rays and high-energy neutrinos.

The \textit{Vera C. Rubin} Observatory (hereafter \textit{Rubin}) and its Legacy Survey of Space and Time (LSST) is poised to revolutionize transient astronomy with its unprecedented depth and cadence. \textit{Rubin} is slated to begin operations and have first light in 2025\footnote{\url{https://www.lsst.org/about/project-status}}. It will have a wide-field telescope with an 8.4 meter mirror and six photometric filters. LSST will be a 10-year-long survey, and each year its vast data volume will enable the discovery and monitoring of thousands of TDEs \citep{bricman_prospects_2020}, offering a unique opportunity to study their properties and statistical distribution in detail. However, the sheer volume of data and lack of spectroscopy necessitates the development of efficient and accurate classification algorithms to distinguish TDEs from other extragalactic transient events, such as supernovae (SNe) and active galactic nuclei (AGNs).

Many classifiers based on machine learning (ML) algorithms are already in use and play a crucial role in identifying optical transients \citep{villar_superraenn_2020, hosseinzadeh_photometric_2020, sheng_neural_2023}. While there are existing ML classifiers designed to categorize TDEs using real \textit{Zwicky} Transient Facility (ZTF) data and host-galaxy information \citep{stein_texttttdescore_2024, gomez_identifying_2022, sheng_neural_2023}, none have been specifically trained to recognize TDEs using the most recent simulated LSST-like data.

In this paper we introduce a novel photometric classifier specifically tailored to identify TDEs in LSST data using only light-curve information, i.e., without needing spectroscopic confirmation, host-galaxy associations, or redshifts. By leveraging ML techniques on simulated datasets, and using only photometry, our approach aims to significantly improve TDE detection rates and facilitate the construction of robust TDE catalogs for astrophysical and multi-messenger investigations. The code is available on GitHub\footnote{\url{https://github.com/kunalfzu/TDE}}. In Sect.~2 we describe the dataset used for our analysis. Section~3 details the light-curve fitting and feature extraction procedures, which is followed by an examination of feature distributions and correlations in Sect.~4. We present our classifier architecture and report its performance metrics in Sect.~5. Finally, Sect.~6 discusses the implications of our results, and Sect.~7 offers concluding remarks.

\section{Dataset}
\label{sec:dataset}

In anticipation of the \textit{Rubin} LSST, several large-scale simulation efforts (for, e.g., CosmoDC2; \citealt{korytov_cosmodc2_2019}) have been initiated to provide realistic light curves for a variety of transient and variable objects. These simulations aim to mirror LSST’s planned cadence, photometric depths, and noise characteristics, thereby allowing researchers to develop and test ML classifiers under conditions that closely approximate future survey data.

One of the most comprehensive such efforts is the Extended LSST Astronomical Time Series Classification Challenge (ELAsTiCC), organized by the LSST Dark Energy Science Collaboration (DESC)\footnote{\url{https://portal.nersc.gov/cfs/lsst/DESC_TD_PUBLIC/ELASTICC/}}. The primary goal of ELAsTiCC is to create synthetic datasets with accurate astrophysical diversity and realistic observational effects, enabling the community to develop, benchmark, and refine algorithms for transient classification well in advance of LSST operations. Building on the success of the Photometric LSST Astronomical Time-series Classification Challenge (PLAsTiCC; \citealt{kessler_models_2019}), ELAsTiCC incorporates updated survey strategies, enhanced modeling of various transient populations, realistic host galaxy associations and probabilistic photometric redshifts.

For this work, we focused on the second release of ELAsTiCC (dubbed ELAsTiCC2), which provides an even more advanced simulation of extragalactic transients. In Sect.~\ref{sec:elasticc2} we outline how we extracted and preprocessed the subset of ELAsTiCC2 data relevant for our TDE classifier, along with the specific transient classes that pose the greatest risk of contaminating a TDE candidate sample. 

\subsection{ELAsTiCC2}
\label{sec:elasticc2}

ELAsTiCC2 is a simulation of transients observed by LSST under realistic observing conditions. ELAsTiCC2 uses a more current baseline 3.2 LSST cadence over 3 years, including a rolling cadence in years 2--3 and including deep drilling fields. For ELAsTiCC2, light curves for 4.1~million transient and variable objects were simulated, yielding 62~million detections and 990~million forced photometry points. The ELAsTiCC2 photometry provided is as observed; we did not de-redden the light curves. For deployment on real LSST data, we plan to mitigate Galactic contamination using: (i) a \textit{Gaia} parallax and/or proper–motion veto; (ii) a nuclear–offset prior once a host is associated; (iii) a Galactic–latitude mask (e.g., exclude $\lvert b\rvert\lesssim10^{\circ}\text{--}15^{\circ}$); and (iv) simple variability–timescale rules to suppress rapid stellar variables. In addition, \textit{Rubin} LSST Data Release~1 (DR1) will provide photometric redshifts and star–galaxy separation scores that can be used as early checks before our Gaussian process+XGBoost model. To train and test our classifier, in addition to the TDEs themselves, we used only the extragalactic models (and subclasses within) that are most likely to be TDE impostors: AGNs, superluminous supernovae (SLSNe), and SNe of various types (Ia, Iax, Ib, Ic, Ic-BL, IIb, and IIn).

\subsection{TDE simulation and modeling}
\label{sec:tde_simulation}

ELAsTiCC2 employs the Modular Open Source Fitter for Transients (\texttt{MOSFiT}) TDE model to generate realistic light curves. These models are based on hydrodynamical simulations from \cite{guillochon_hydrodynamical_2013} and \cite{mockler_weighing_2019}, which simulate the disruption of stars by SMBHs. Using 11 real TDE light curves as a basis, the \texttt{MOSFiT} model parameters were fitted to generate 745 spectral energy distributions. Using these spectral energy distributions, thousands of TDE light curves were created with the SuperNova ANAlysis (SNANA) software \citep{kessler_snana_2009}, ensuring a diverse and comprehensive training set. A more detailed description of the model of sources is given in \cite{kessler_models_2019} and references therein. The simulated TDEs incorporate known volumetric rates and redshift dependences to reflect the expected distribution in the universe. However, the model is limited by the relatively low number of fitted TDEs, which may result in less accurate representation of diversity and demographics within the TDE population. Notably, newer TDEs found in literature, such as those with double peaks, pre-peak bumps, late-time plateaus, or re-brightening, are not included in the simulated TDE model in ELAsTiCC2.

\section{Light-curve fitting and feature extraction}
\label{sec:LC_fitting}

\subsection{Light-curve quality cut}
\label{sec:quality_cut}

\begin{figure}[htbp]
    \centering
    \includegraphics[width=0.9\linewidth]{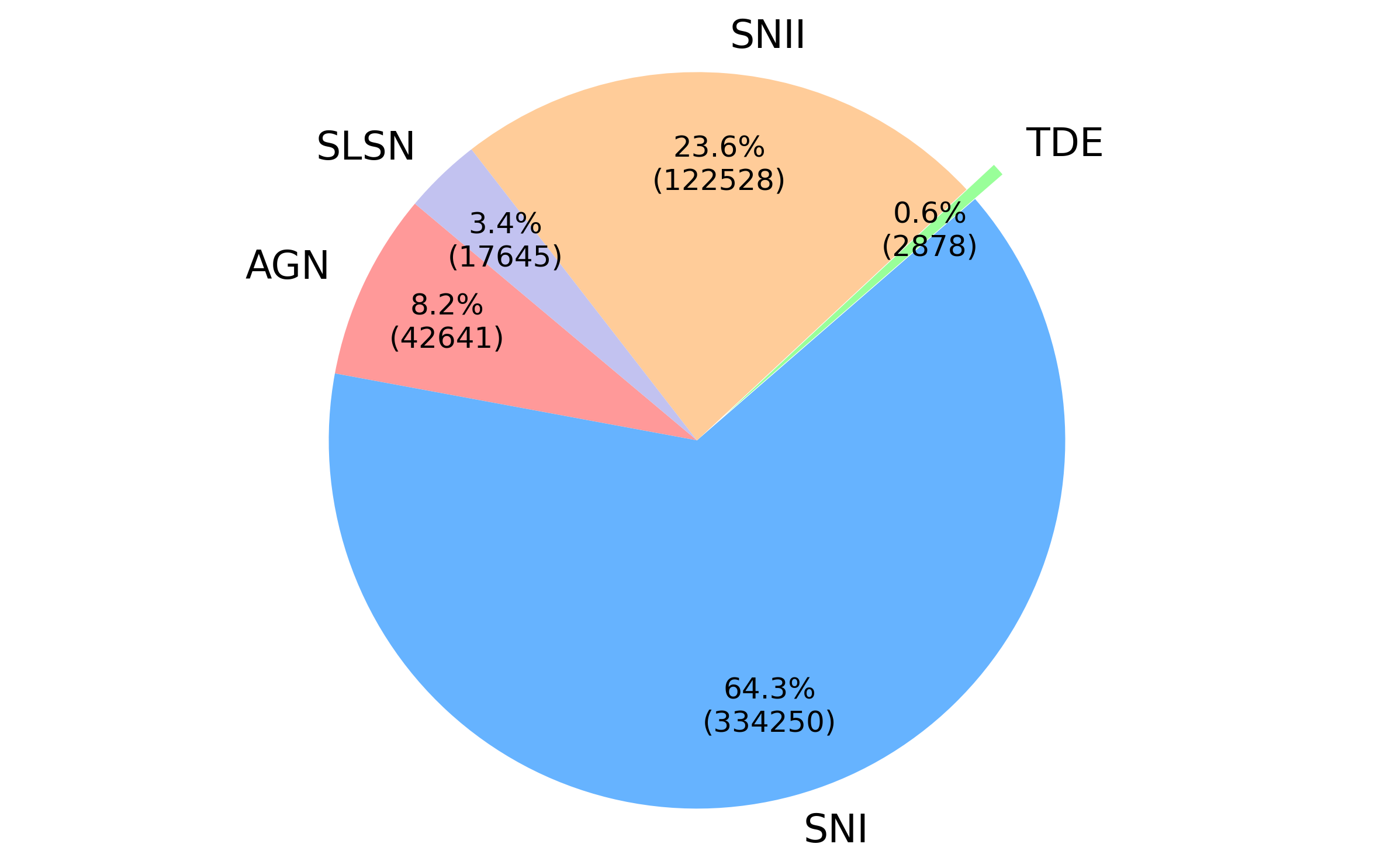}
    \caption{Sample population after the light-curve quality cut. Of the remaining objects, 2878 (0.6\%) are TDEs.}
    \label{fig:pie_chart}
\end{figure}

To ensure a bare minimum data quality, we required that each light curve have at least three observations in any photometric band with a signal-to-noise ratio greater than 5 and a calibrated flux exceeding 100, corresponding to approximately 22.5~mag. These cuts help filter out noisy and unreliable light curves, thereby enhancing the robustness of the classifier. The initial and quality-cut counts and the population of the remaining sources are shown in Table~\ref{tab:class_counts} and Fig.~\ref{fig:pie_chart}, respectively.

\begin{table}[htbp]
\centering
\caption{Objects before and after quality cuts and the retention fraction.}
\label{tab:class_counts}
\begin{tabular}{lrrr}
\toprule
\textbf{Class} & \textbf{Initial count} & \textbf{Quality cut} & \textbf{\% kept} \\
\midrule
AGN   & 108556   & 42641   & 39.3\% \\
SLSN  & 32760    & 17645   & 53.9\% \\
SNI   & 1968937  & 334250  & 17.0\% \\
SNII  & 1172811  & 122528  & 10.4\% \\
TDE   & 13076    & 2878    & 22.0\% \\
\bottomrule
\end{tabular}
\tablefoot{The percentage indicates the fraction of objects retained after the cut.}
\end{table}

\subsection{Fitting light curves with Gaussian processes}
\label{sec:GPs}
Gaussian processes (GPs) provide a flexible and probabilistic approach to modeling light curves \citep[see][]{aigrain_gaussian_2023}. They are nonparametric and, hence, can be trained also on exotic TDEs (those exhibiting double peaks, pre-peak bumps, late-time plateaus, re-brightening, etc.) and other potential subtypes that may not have been discovered yet. By fitting GPs to the photometric data, we can extract features such as variability timescales, color, rise and fade times, and color evolution. These features can then serve as inputs to the classifier, enabling it to distinguish TDEs from other transient phenomena based on their unique photometric signatures. Our implementation of GP follows \cite{boone_avocado_2019}. We began by fitting the light curves of all object types with GP using the \texttt{George} package \citep{ambikasaran_fast_2015}. After experimenting with several kernels available in \texttt{George}, we settled on a 2D Matérn-3/2 kernel (initialized with 100~days and 6000~\AA) for amplitude (flux), time length scale, and wavelength length scale parameters on a per-object basis. While no explicit bounds for the hyperparameters were enforced, we used an optimizer that retries up to 3 times with small perturbations and chooses the best result. This kernel proved effective for fitting TDE-like light curves. We used the central values of each LSST band as effective wavelengths and optimized the fit via the \texttt{minimize} function from \texttt{scipy}, which uses the Broyden–Fletcher–Goldfarb–Shanno optimization algorithm. Using a kernel in both time and wavelength allows the GP to leverage cross-band information to interpolate a light curve even when it is poorly sampled in any given band. An example of a fit is shown in Fig.~\ref{fig:}, where the \(u\) band has no observation before the peak, yet the GP, using cross-band information, predicts the light curve reasonably well.

\subsection{Feature extraction}
\label{sec:feature_extraction}
Tidal disruption events in the literature have certain known characteristics that help distinguish them from other transients such as AGNs and SNe. These include their bluer color, minimal color evolution, and longer rise and fade timescales. Hence, we extracted the following features:
\begin{itemize}[label=\textbullet, itemsep=0.75em]
    \item Rise and fade timescale: Drop of 1~mag on either side of the GP-estimated \(g\)-peak.
    \item Weighted mean color \(g-r\) and \(r-i\): Minimum of one observation within rise and fade times for pre- and post-peak color, respectively (using \(g\)-peak for \(g-r\), \(r\)-peak for \(r-i\)).
    \item Weighted color evolution: Slope of a fitted line to at least two observations in \(g-r\) and \(r-i\) within the rise and fade times, for pre- and post-peak color, respectively.
    \item GP hyperparameters: Amplitude, length scale in time, and length scale in wavelength.
    
\end{itemize}

\begin{figure}[htbp]
    \centering
    \includegraphics[width=1\linewidth]{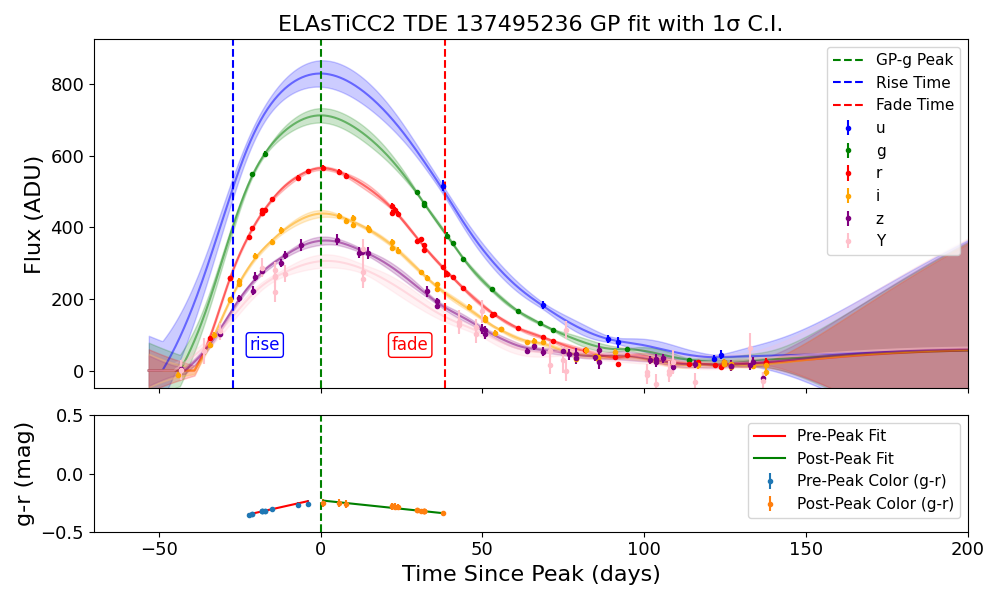}
    \caption{Example light curve with a GP fit.\ We show the rise and fade times as well as the pre- and post-peak \(g-r\) color within the rise and fade times.}
    \label{fig:}
\end{figure}

Although all six LSST bands are fit with GP, we focused on the three most frequently observed bands (\(g\), \(r\), and \(i\)) for feature extraction. First, to empirically estimate timescales, we identified the \(g\)- and \(r\)-band peaks based on the GP-estimated maximum flux, then measured the rise and fade timescales in the \(g\) band, where brightness decreases by one magnitude (\(\approx 40\%\) of the original flux) on either side of the peak. While this approach is conceptually similar to \cite{hammerstein_final_2022} our implementation relies on GP interpolation rather than fitting a line between just two adjacent points. Choosing 1 magnitude to measure the timescales proved to be a robust and stable method as the part around the peak of the light curve is usually better defined than near the baseline. Second, using the \(g\)-peak for \(g-r\) and the \(r\)-peak for \(r-i\), we computed pre- and post-peak \(g-r\) (and \(r-i\)) within the corresponding rise and fade times, where at least one observation in either \(g\) or \(r\) (and \(r\) or \(i\)) is present. For instance, if there is an observation in \(g\) but not \(r\), the GP curve provides an interpolated flux in \(r\). Figure~\ref{fig:} illustrates this scenario, where an observation in \(r\) exists near the peak, but none in \(g\); nonetheless, we still determined \(g-r\) via the GP prediction. Third, to quantify color evolution, we fit a line to the \(g-r\) (and \(r-i\)) measurements (requiring at least two points) for both the rise and fade intervals, weighting each data point by its uncertainty. Finally, we extracted the three GP hyperparameters — amplitude, time, and wavelength length scales — leading to a total of 13 features per light curve. The GP fitting procedure and feature extraction is relatively swift in practice and takes about 0.2~seconds per object with parallelization on a 16-core CPU. 

\subsection{Success rate}

\begin{figure}[htbp]
    \centering
    \includegraphics[width=1\linewidth]{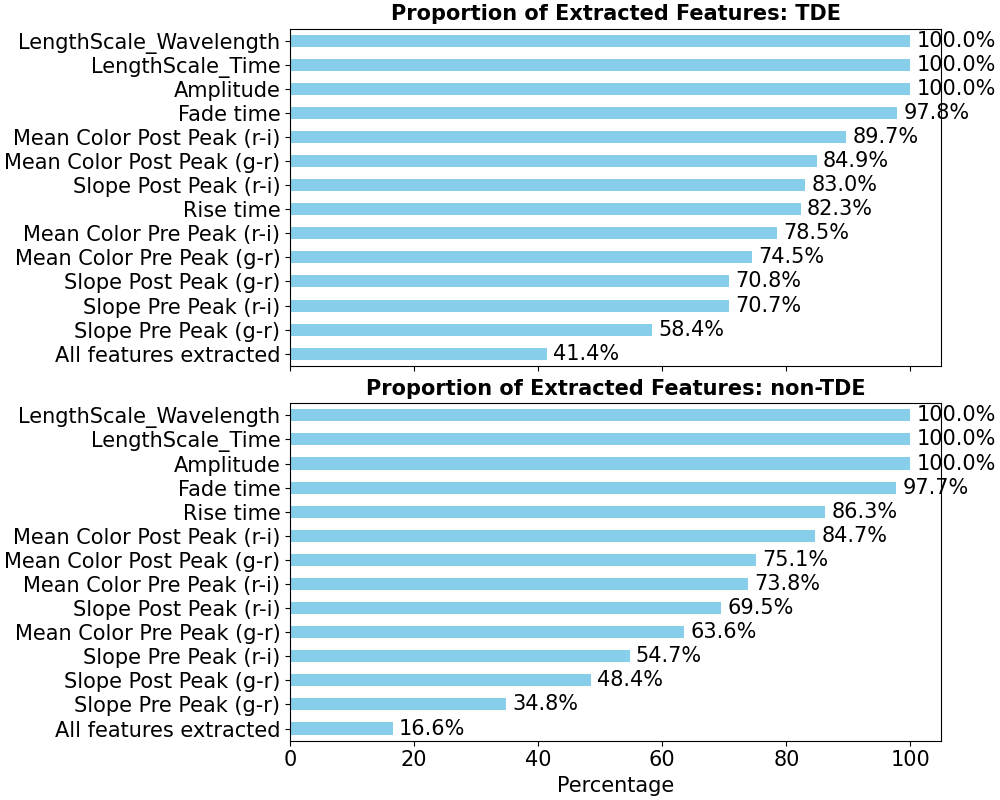}
    \caption{Success rate of feature extraction for TDEs and non-TDEs.}
    \label{fig:successrate}
\end{figure}

Up to this point, we did not discard any objects. After fitting all objects of various types that passed the light-curve quality cut, we calculated the success rate of extracted features, as shown in Fig.~\ref{fig:successrate}. As expected, the GP hyperparameters are successfully calculated for all objects. For TDEs, the fade time was computed for nearly all objects, largely because we extended the GP predictions manually after the peak. About 40\% of TDEs had all features successfully extracted. For the remaining TDEs, some or all of the features could not be extracted, predominantly pre-peak features and slopes, due to low number of data points (a minimum of two points are needed within the rise and fade timescales, as detailed in Sect.~\ref{sec:feature_extraction}).

\section{Feature distributions and correlations}
\label{sec:features_correlations}

The extracted features provide strong discriminatory power for classifying TDEs against other extragalactic transients such as SNe, AGNs, and SLSNe. We focused on color metrics (mean post-peak \texttt{g-r} and \texttt{r-i}), color slopes (rate of \texttt{(g-r)} or \texttt{(r-i)} change), and GP hyperparameters \texttt{LengthScale\_Time} and \texttt{LengthScale\_Wavelength}, along with rise and fade times. This section highlights a few illustrative 2D projections and a correlation matrix. Histograms of all extracted features are given in Appendix~A.

\begin{figure}[htbp]
    \centering
    \includegraphics[width=0.48\textwidth]{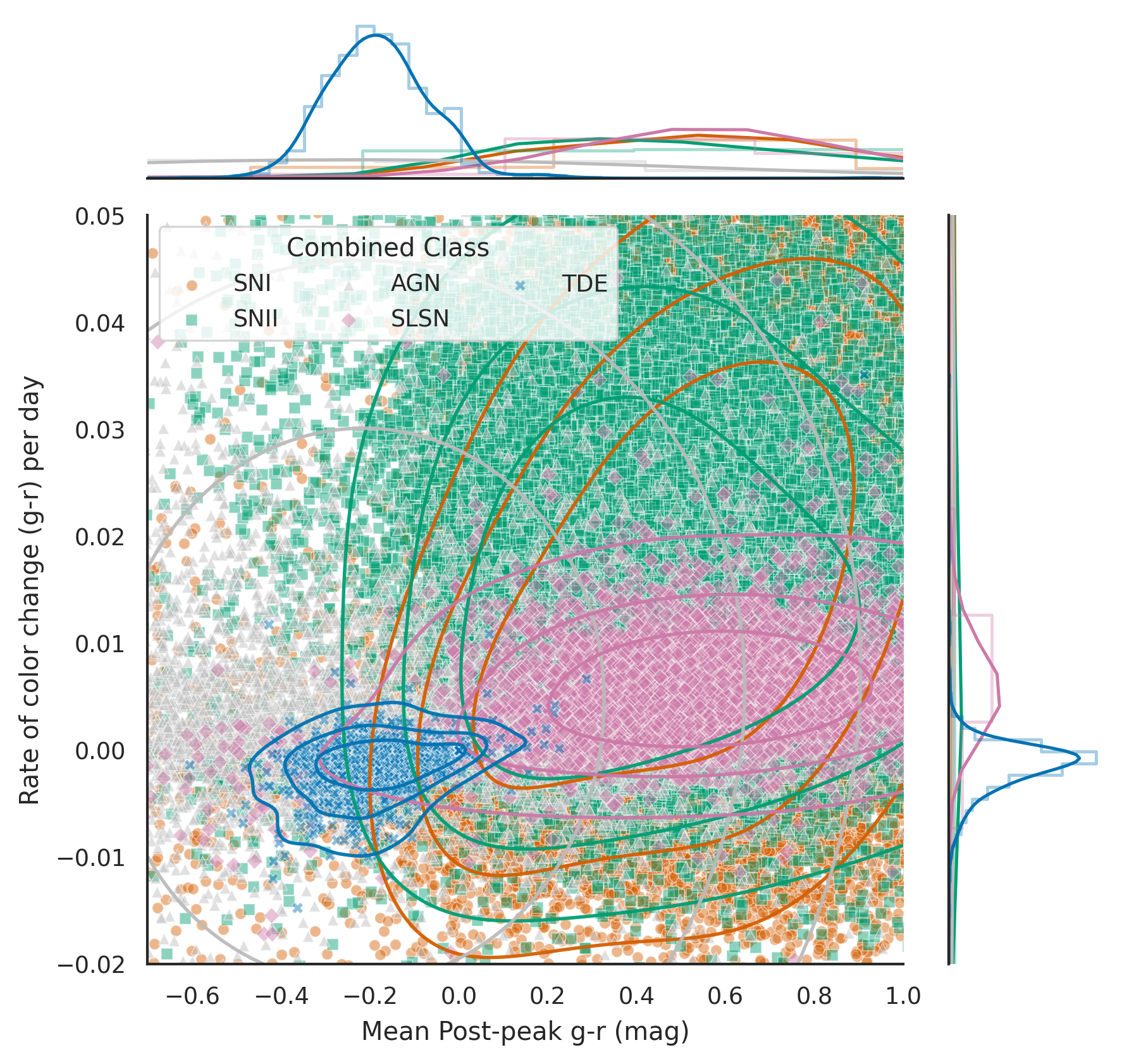}
    \caption{Scatter plot of the mean post-peak \texttt{g-r} (horizontal axis) vs. its rate of change (vertical axis), illustrating how TDEs (blue crosses) tend to cluster around smaller color values and lower color evolution rates, in contrast to other transient classes. Contours denote the 50\%, 80\%, and 95\% highest-density regions of the per-class 2D kernel density estimate.}
    \label{fig:color_slope_vs_color}
\end{figure}

\begin{figure}[htbp]
    \centering
    \includegraphics[width=0.52\textwidth]{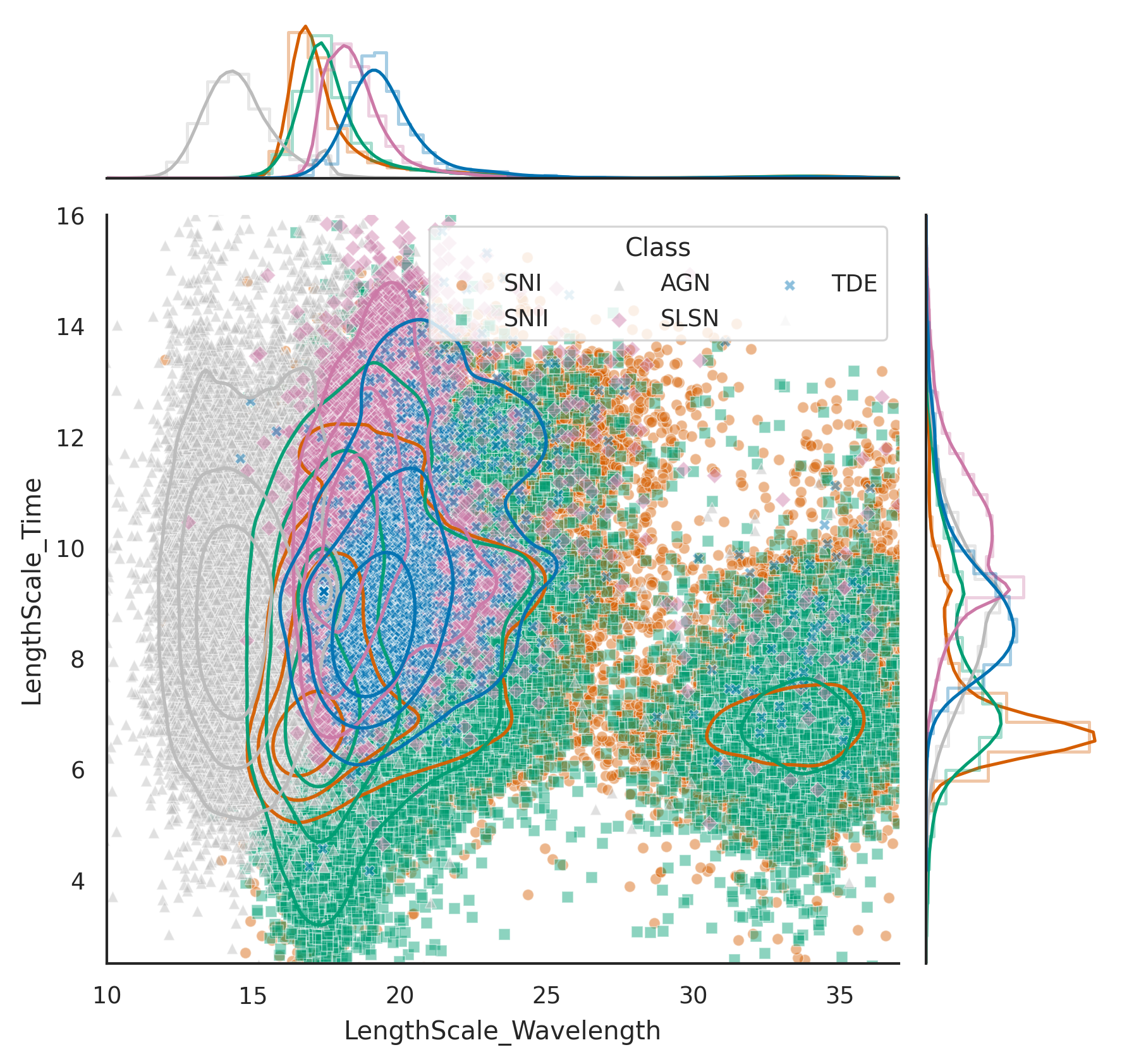}
    \caption{2D distribution of the GP hyperparameters \texttt{LengthScale\_Time} (vertical axis) vs. \texttt{LengthScale\_Wavelength} (horizontal axis), showing how each transient class occupies distinct regions in this parameter space. TDEs (blue crosses) and SLSNe (pink diamonds) typically have moderate \texttt{LengthScale\_Time} and \texttt{LengthScale\_Wavelength} values, reflecting narrower spectral and temporal variations compared to SNe and AGNs. Contours denote the 50\%, 80\%, and 95\% highest-density regions of the per-class 2D kernel density estimate.}
    \label{fig:lengthscale_time_wavelength}
\end{figure}

\begin{figure}[htbp]
    \centering
    \includegraphics[width=0.48\textwidth]{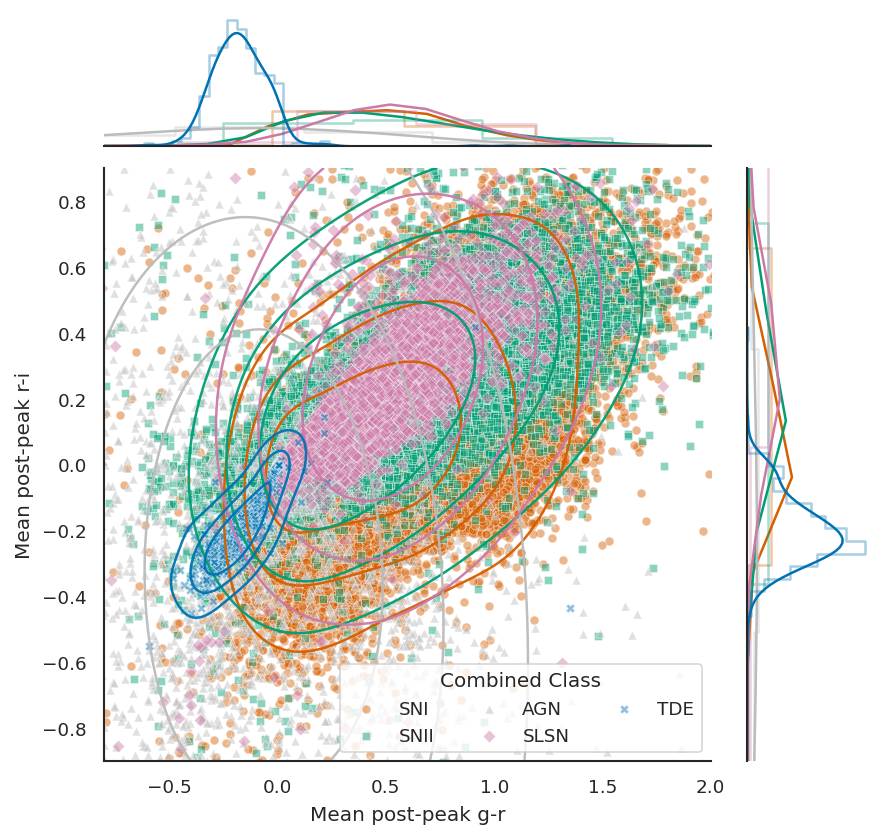}
    \caption{Scatter plot of mean post-peak \texttt{g-r} (horizontal) vs.\ mean post-peak \texttt{r-i} (vertical), with 1D distributions (histograms) shown along the top and right axes. TDEs (blue crosses) appear bluer than SNe (orange circles and green squares) and SLSNe (pink diamonds), while AGNs (gray triangles) span a broader range of colors. Contours denote the 50\%, 80\%, and 95\% highest-density regions of the per-class 2D kernel density estimate.}
    \label{fig:color_vs_color}
\end{figure}

\begin{figure}[htbp]
    \centering
    \includegraphics[width=0.48\textwidth]{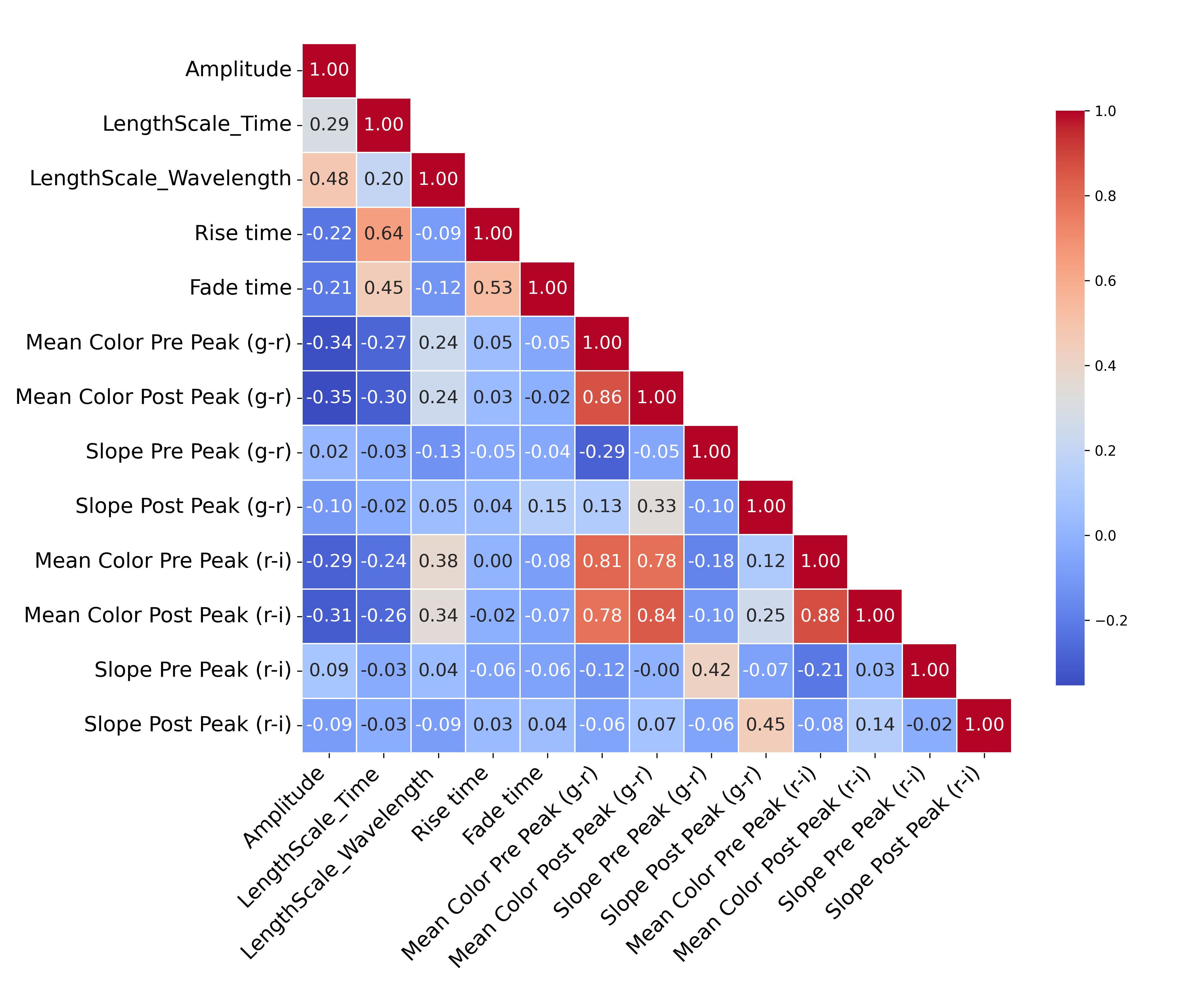}
    \caption{Correlation matrix of the selected features, indicating pairwise Spearman or Pearson correlation coefficients. Warmer colors correspond to higher positive correlations, while cooler colors indicate negative correlations. Notably, \texttt{Mean\_Color\_Post\_Peak\_gr} and \texttt{Mean\_Color\_Post\_Peak\_ri} show a strong positive correlation, reflecting common color-evolution behavior in TDEs.}
    \label{fig:correlation_matrix}
\end{figure}

Figure~\ref{fig:color_slope_vs_color} shows that TDEs (blue points) have smaller mean post-peak \texttt{g-r} values and lower \texttt{(g-r)} color-evolution rates compared to other transient classes. AGNs (yellow) often cluster around near-zero evolution but vary widely in mean \texttt{g-r}, while SNe (red for SNe I and green for SNe II) are scattered across a broader range of \texttt{g-r} color. The broad clump of SLSNe (purple) indicates more extreme color and evolution values.

In Fig.~\ref{fig:lengthscale_time_wavelength} we plot two key GP hyperparameters: \texttt{LengthScale\_Time} and \texttt{LengthScale\_Wavelength}. TDEs typically occupy intermediate values for both parameters, aligning with their characteristic timescales and gradual color changes. By contrast, some SN subtypes can display distinctly larger (or smaller) \texttt{LengthScale\_Time} due to rapid declines post-peak or faster color evolution. AGNs particularly stand out, as they are known to exhibit high variability both in time and wavelength.

Figure~\ref{fig:color_vs_color} illustrates a color--color diagram, showing the relationship between mean post-peak \texttt{g-r} and \texttt{r-i}, along with marginal distributions projected on the top and right axes. TDEs cluster toward bluer \texttt{g-r} and \texttt{r-i}, while AGNs exhibit extended tails in \texttt{g-r}. SNe I and SNe II partially overlap in color space but deviate distinctly from TDE-like blue colors.

Finally, Fig.~\ref{fig:correlation_matrix} presents a correlation matrix for the main features. Key observations include the strong interdependence of color features, reflecting the overall spectral evolution of the transients. Likewise, rise and fade times exhibit moderate correlations with the GP \texttt{LengthScale\_Time}, consistent with physically longer timescales leading to broader or slower light-curve variations.

Collectively, these 2D projections and the correlation matrix highlight the photometric diversity among transients, demonstrating how our selected features help separate TDEs from other classes in the parameter space used for classification.

\section{Classifier architecture and performance}
\label{sec:classifier_architecture}

Having processed the light curves of all object types, we fed the extracted features into a ML classifier, including XGBoost \citep{chen_xgboost_2016} and random forest from the scikit-learn Python package \citep{pedregosa_scikit-learn_2011}.
Since we had enough TDEs (2878 TDEs, 517064 non-TDEs), we could split the dataset evenly into a training set and a test set without the need for data augmentation, even though TDEs constitute only 0.6\% of the total. With 1439 TDEs and 258532 non-TDEs in each set, we trained and test our ML classifier. First, we prepared data for the XGBoost model. Any missing data in our 13 features were replaced with a placeholder (e.g., \(-999\)) or the mean across the dataset for a given feature. We find that neither choice significantly impacts the classification. Next, to optimize XGBoost, we defined a hyperparameter search space, outlining reasonable ranges and distributions for key parameters (e.g., number of trees, learning rate, and maximum depth). This search space captures the most influential settings in XGBoost, allowing the model to explore configurations that may enhance performance.

Because achieving high precision (or purity) is central to our science goals (i.e., constructing an optical TDE candidate catalog for statistical studies of correlations with other messengers, such as high-energy neutrinos), we set a target precision of 95\%. To integrate this requirement, we implemented a custom scoring function that identifies solutions meeting or exceeding this precision threshold, then selects among them the one yielding the highest recall (or completeness). This encourages the classifier to minimize false positives while recovering as many true positives as possible. We performed the hyperparameter search with stratified \(k\)-fold cross-validation and a randomized search approach using \texttt{RandomizedSearchCV} from \texttt{scikit-learn}. Stratified splitting ensures each fold has a similar class distribution, enabling fairer comparisons of different settings. Randomized search accelerates tuning by sampling diverse configurations within the predefined ranges of key parameters, striking a balance between computational efficiency and thoroughness. The resulting hyperparameter values are shown in Table~\ref{tab:xgb_params}.

\begin{figure}[htbp]
    \centering
    \includegraphics[width=0.8\linewidth]{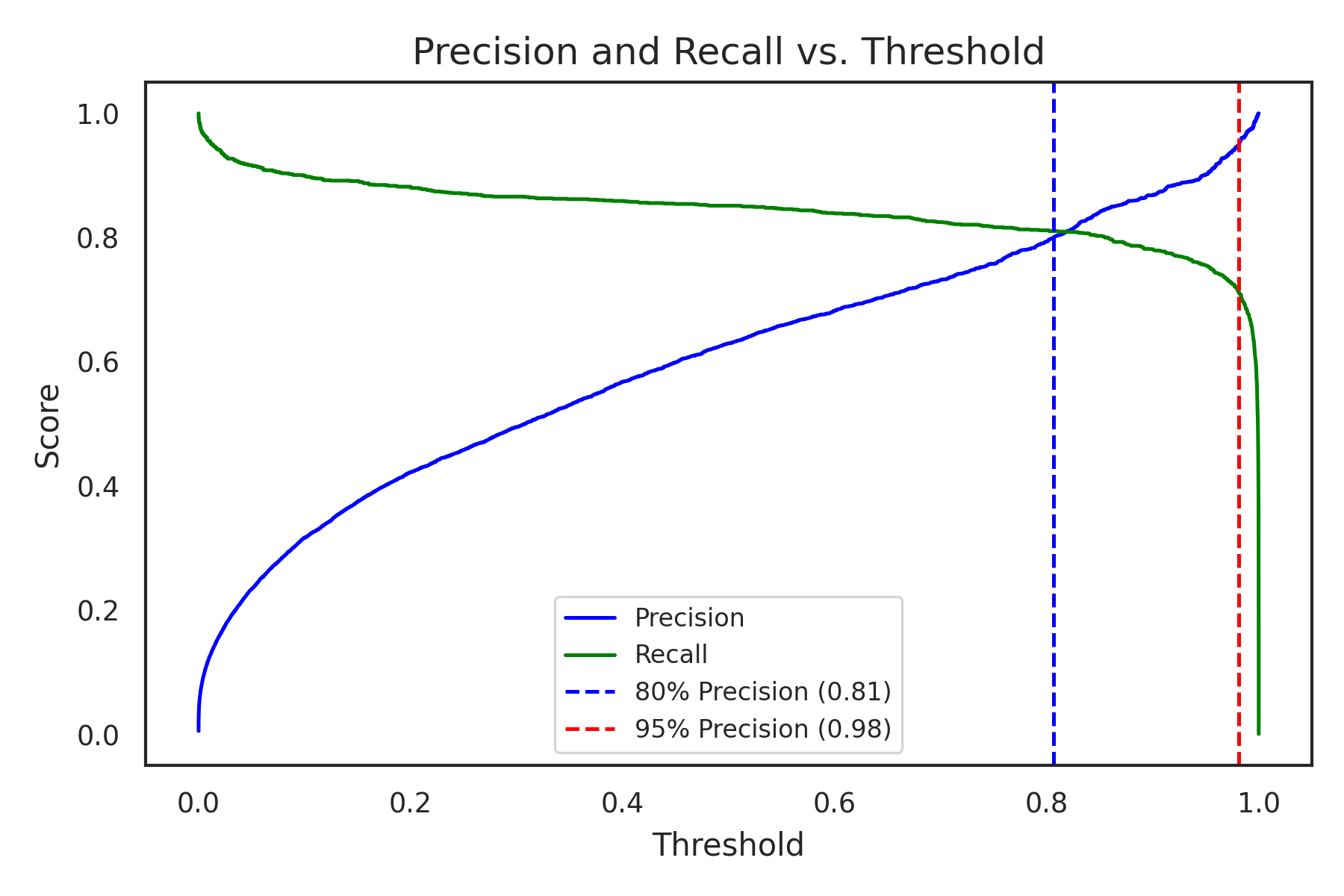}
    \caption{Precision and recall as a function of classifier threshold.}
    \label{fig:xgb_precision_recall}
\end{figure}

After identifying the best hyperparameters, we applied the optimized XGBoost model to the test set to generate decision scores for the positive (TDE) class. Inspired by \citet{stein_texttttdescore_2024}, we converted these decision scores into binary class labels by defining decision thresholds for three scenarios: 80\% precision (balanced), 95\% precision (high-purity), and 95\% recall (high-completeness). Figure~\ref{fig:xgb_precision_recall} shows how precision and recall vary with the threshold.

In the balanced case (at least 80\% precision), the classifier recovers 81\% (1166) of TDEs while correctly classifying 99.9\% (258242) of non-TDEs, with only 0.1\% (291) misclassified as TDEs.

\begin{table*}[htbp]
\centering
\caption{Optimized hyperparameter values for the XGBoost model.}
\label{tab:xgb_params}
\begin{tabular}{l l p{12.5cm}}
\toprule
\textbf{Hyperparameter} & \textbf{Value} & \textbf{Description} \\
\midrule
\texttt{colsample\_bytree}  & 0.86 & Fraction of features (columns) used per tree. Helps prevent overfitting by limiting how many features each tree can rely on. \\
\texttt{gamma}              & 3.3 & Minimum loss reduction required to make a further partition on a leaf node. Higher values help control complexity by preventing excessive splitting. \\
\texttt{learning\_rate}     & 0.09 & Shrinkage factor that reduces each tree’s contribution. Lower rates often improve accuracy but require more trees. \\
\texttt{max\_depth}         & 7    & Maximum number of levels in each tree. Deeper trees can learn complex patterns but risk overfitting. \\
\texttt{min\_child\_weight} & 7    & Minimum sum of instance weights needed in a child (leaf). Larger values restrict the model from creating leaves with few samples. \\
\texttt{n\_estimators}      & 291  & Number of boosting rounds (trees). More rounds can improve performance but also increase overfitting risk. \\
\texttt{scale\_pos\_weight} & 208.8 & Balance for class imbalance. Higher values make the model pay more attention to the minority (positive) class. \\
\texttt{subsample}          & 0.8 & Fraction of training data sampled per tree. Lower fractions help reduce correlation between trees but can skip important examples. \\
\bottomrule
\end{tabular}
\end{table*}

\begin{figure*}[htbp]
    \centering
    \includegraphics[width=2\columnwidth]{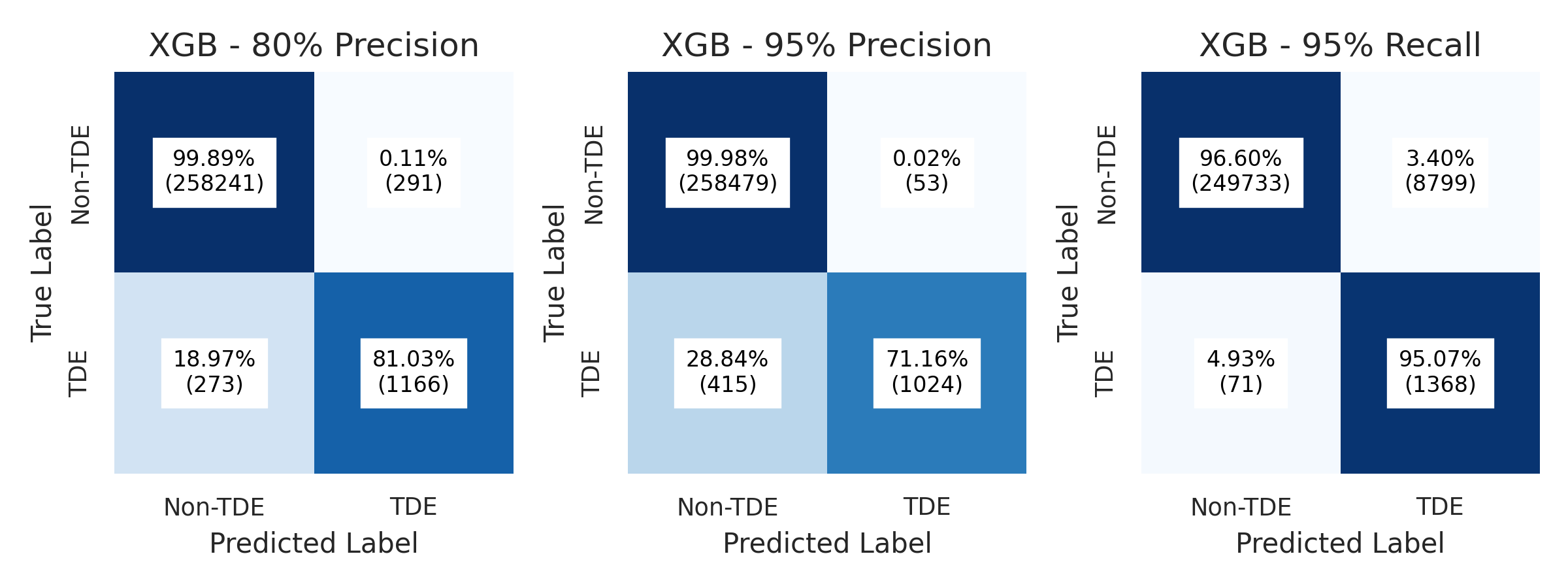}
    \caption{XGB truth-normalized confusion matrices for different thresholds. True classes are on the vertical axis, and predicted classes on the horizontal axis.}
    \label{fig:xgb_confusion_matrix}
\end{figure*}

\begin{figure}[htbp]
    \centering
    \includegraphics[width=1\linewidth]{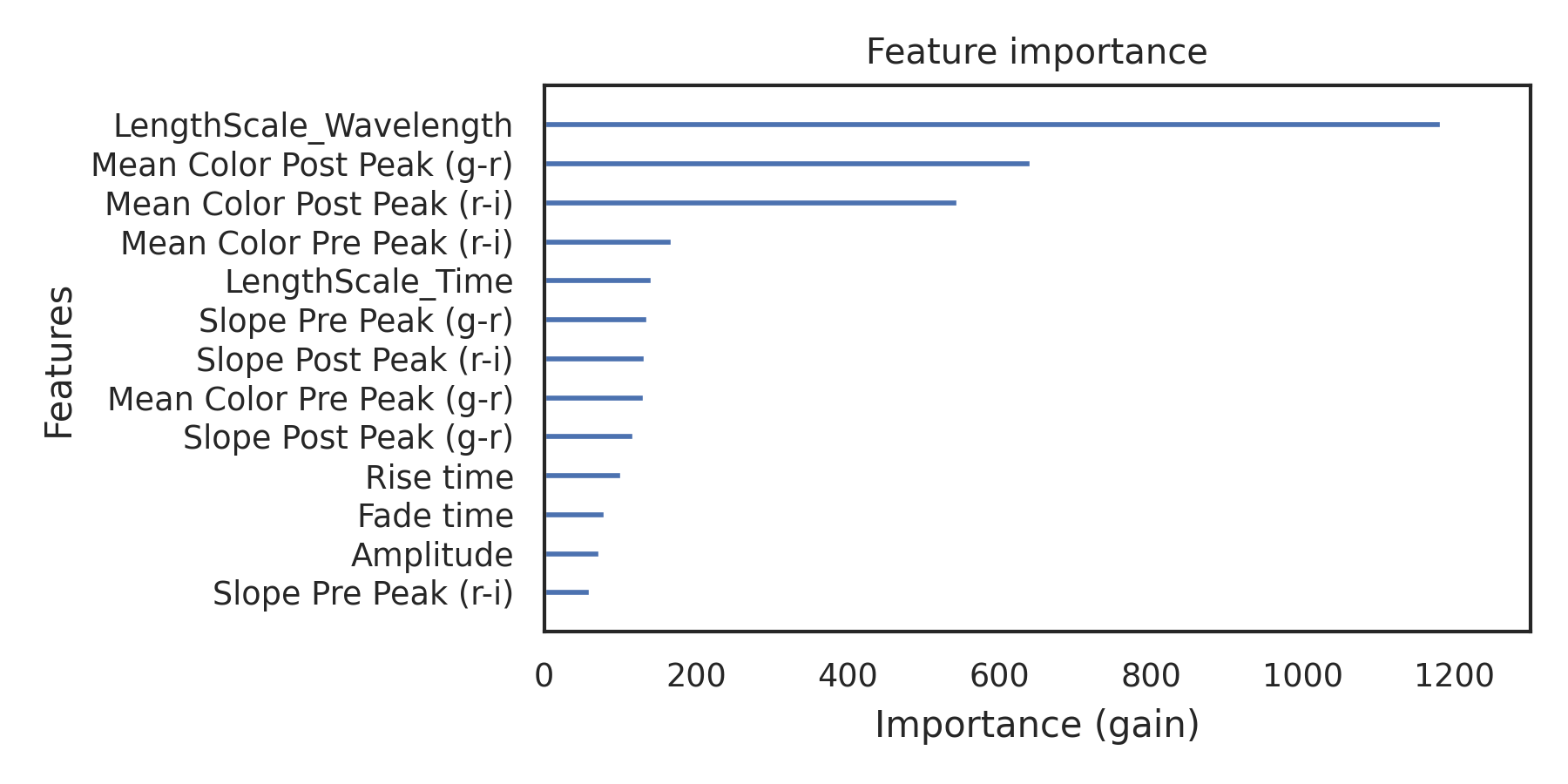}
    \caption{Relative importance of extracted features calculated by XGB using the gain type.}
    \label{fig:xgb_feature_importance}
\end{figure}

\section{Discussion}
\label{sec:discussion_conclusion}

The results presented in this work highlight both the efficacy and flexibility of our photometric classifier for TDEs. By leveraging GP for feature extraction and XGBoost for classification, we achieved high precision and recall on simulated LSST-like data (ELAsTiCC2). Below, we summarize and discuss the key findings, referencing relevant figures that illustrate our conclusions.

\subsection{Feature importance and classifier insights}
Figure~\ref{fig:xgb_feature_importance} provides a visual summary of feature importance as measured by the XGBoost gain metric. Several observations stand out:
\begin{itemize}
    \item Color-based features dominate: The post-peak colors \texttt{(g-r)} and \texttt{(r-i)} rank at the top, indicating their strong discriminating power. TDEs are known to remain bluer than SNe or AGNs, particularly in the post-peak phase.
    \item GP hyperparameters are influential: Features like \texttt{LengthScale\_Wavelength} and \texttt{LengthScale\_Time} highlight how characteristic timescales and wavelength behaviors effectively differentiate TDEs from other transients.
    \item Color evolution matters: The slope post peak \texttt{(g-r)/(r-i)} distinguishes TDEs from other transients, as the rate of color change per day is minimal for TDEs, especially after peak.
    \item Time-based features are moderate in rank: While rise time and fade time do contribute, the classifier places more emphasis on color information and the GP length scales.
    \item Amplitude is less critical: Brightness alone, without color or temporal information, proves insufficient for robustly distinguishing TDEs from other transients, hence its comparatively lower feature importance.
\end{itemize}

\subsection{Threshold tuning and the precision--recall trade-off}
We employed a threshold-based classification approach that allows users to select operating points that match their scientific objectives. Figure~\ref{fig:xgb_precision_recall} shows the precision--recall curve as a function of threshold, while Fig.~\ref{fig:xgb_confusion_matrix} displays confusion matrices for three specific thresholds:
\begin{itemize}
    \item 80\% precision: At this threshold (\(\approx 0.81\)), we obtain about 81\% recall, correctly identifying a substantial fraction of TDEs while only 0.1\% of non-TDEs are misclassified.
    \item 95\% precision: Increasing the threshold to \(\approx 0.98\) yields nearly zero contamination by non-TDEs, albeit at a reduced recall of 71.2\%. This high-purity setting is suitable for building a TDE catalog with minimal false positives.
    \item 95\% recall: By aiming to recover nearly all TDEs (about 95.1\%), the classifier admits a false-positive rate of 3.4\% among non-TDEs. This is ideal when completeness is paramount, potentially guiding spectroscopic or multiwavelength follow-ups.
\end{itemize}

\subsection{Processing efficiency}
Our nonparametric GP framework offers a valuable balance between flexibility and performance. Each object requires only about 0.2~seconds for GP fitting and feature extraction with a 16-core CPU. This translates to \(\approx 27\) hours of computing time for \(\approx 0.52\)~million transients in our ELAsTiCC2 post-quality-cut sample. Such efficiency is critical for handling large-scale surveys like the\ \textit{Rubin} LSST, which will detect millions of transient events. 

\subsection{Generalization to other datasets and further improvements}
Although our current results rely on simulated data, validating the approach on real photometric observations from the ZTF is a natural next step (Bhardwaj et al., in prep.). Potential discrepancies in real-world noise levels, survey cadences, and systematic effects may require additional refinements. To deal with false positives such as galactic sources, artifacts, or moving objects, we can implement mitigating filters as described in Sect.~\ref{sec:dataset}. Moreover, the future availability of photometric redshifts and improved host-galaxy association in LSST data will allow for further enhancements:
\begin{enumerate}
    \item Kernel specialization: Different GP kernels (or kernel mixtures) specifically tailored to particular TDE subclasses could significantly improve performance.
    \item Augmented feature set: Integrating photometric redshift estimates and host-galaxy properties would likely strengthen the classifier’s ability to discriminate TDEs from lower-redshift SNe or higher-redshift AGNs.
    \item Exotic TDEs: Engineering new features or modifying the existing ones for complex or outlier light-curve shapes can make the methodology robust against the diversity of TDE populations.
\end{enumerate}

\section{Concluding remarks}
In summary, our photometric classifier, underpinned by GP-based feature extraction and boosted decision trees, demonstrates high effectiveness in identifying TDEs across a range of precision and recall targets. Its color-driven feature set provides a powerful tool for distinguishing TDEs from a diverse population of extragalactic transients. With forthcoming refinements for handling exotic TDE subtypes and real survey conditions, this method is well poised for the next generation of optical transient studies; it will likely play a crucial role in multi-messenger astrophysics and in helping us understand SMBH environments through TDE observations.

\begin{figure}[htbp]
    \centering
    \includegraphics[width=1\columnwidth]{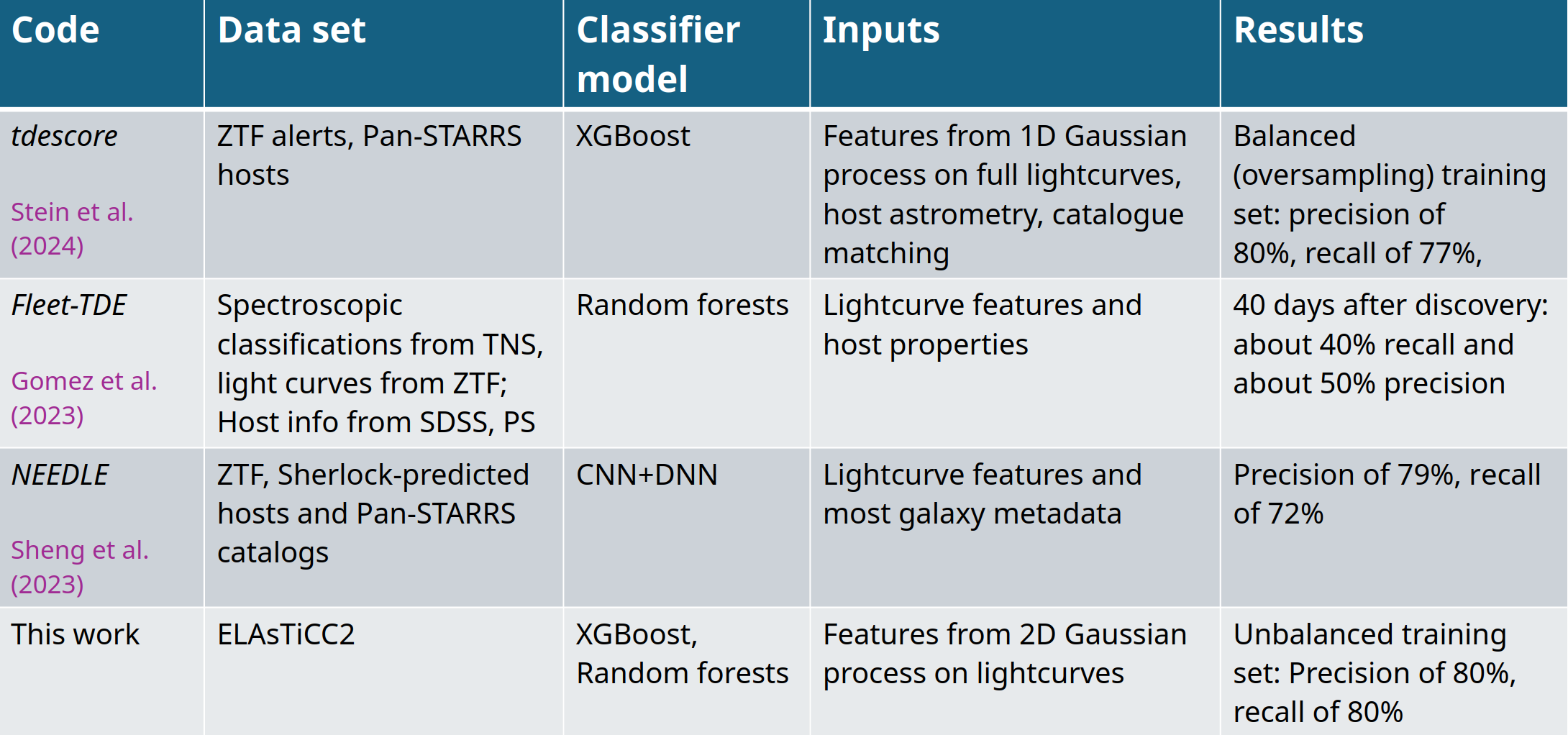}
    \caption{Comparison of this work with dedicated TDE classifiers in the literature.}
    \label{fig:confusion_matrix}
\end{figure}

\begin{acknowledgements}
This work was co-funded by the EU and supported by the Czech Ministry of Education, Youth and Sports (project CZ.02.01.01/00/22\_008/0004632 -- FORTE).
We gratefully acknowledge the support of the Institute of Physics of the Czech Academy of Sciences.
Computations were carried out in part on the HPC cluster Phoebe, operated by the Central European Institute of Cosmology (CEICO) at the Institute of Physics of the Czech Academy of Sciences.
This research made use of the following software packages: \texttt{jupyter}, \texttt{numpy}, \texttt{matplotlib}, \texttt{astropy}, \texttt{sklearn}, \texttt{pandas}, \texttt{seaborn}, \texttt{george}, \texttt{python}, and \texttt{xgboost}.
\end{acknowledgements}

\bibliographystyle{aa}
\bibliography{references}

\clearpage
\begin{appendix}
\onecolumn

\section{Feature distributions}
\label{sec:appendixA}

Here we present the distribution histograms of the extracted features in a single-column format. This layout allows a detailed examination of how various photometric quantities, such as color and amplitude, are distributed across different classes of objects. By comparing these distributions, we can identify patterns that differentiate TDEs from other extragalactic transients.

\begin{figure}[htbp]
    \centering
    \includegraphics[width=\linewidth]{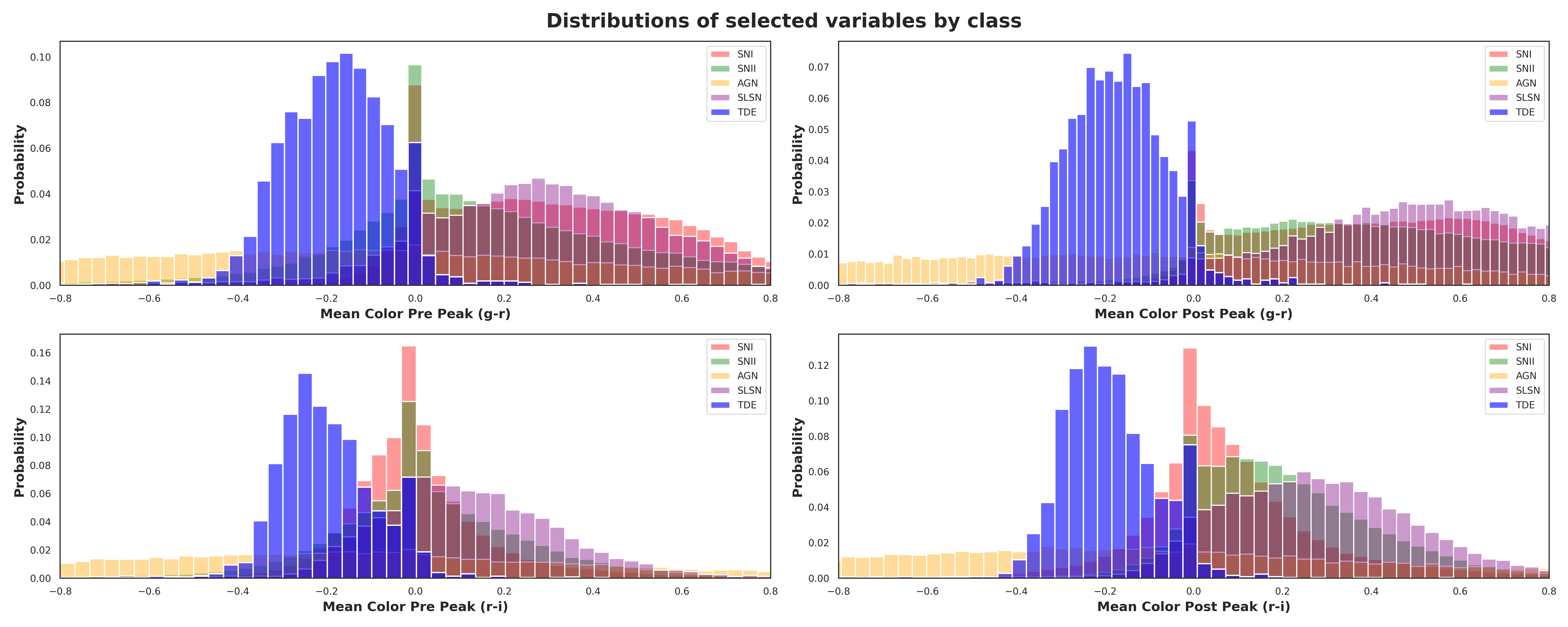}
    \caption{The visible peak at mean color 0.0 in all the plots is a direct consequence of the GP fit failing (and all predicted light-curve bands converging to one curve) due to noisy or poor-cadence light curves.}
    \label{fig:feature_distributions}
\end{figure}

\begin{figure}[htbp]
    \centering
    \includegraphics[width=\linewidth]{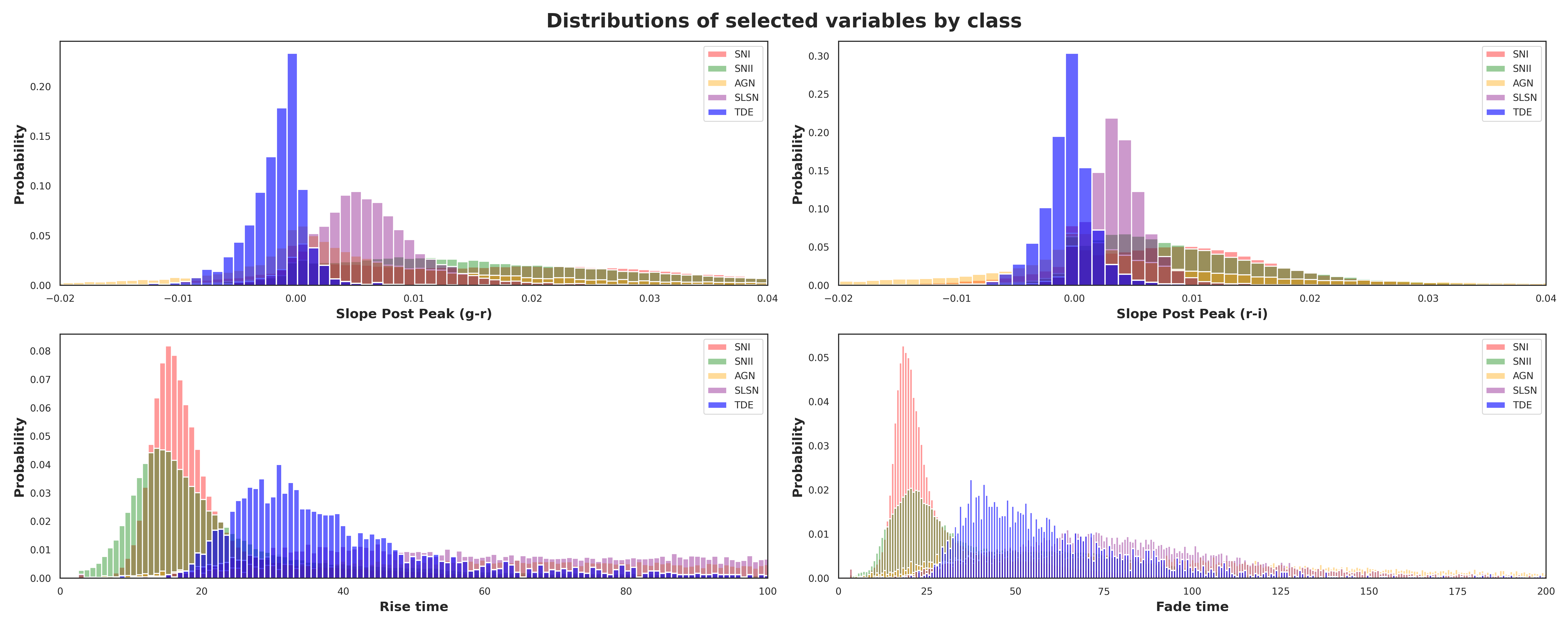}
    \caption{Feature distributions for objects that pass the light-curve quality cut.}
    \label{fig:feature_distributions1}
\end{figure}

\clearpage

\section{Random forest results}
\label{sec:appendixB}

In addition to XGBoost, we employed a random forest classifier, included in the scikit-learn package \citep{pedregosa_scikit-learn_2011}, to compare performance and feature importance across different ensemble methods. The method is similar to that described for XGBoost, with a focus on growing multiple decision trees while randomly sampling both training instances and features to reduce overfitting.

\begin{figure}[htbp]
    \centering
    \includegraphics[width=0.45\linewidth]{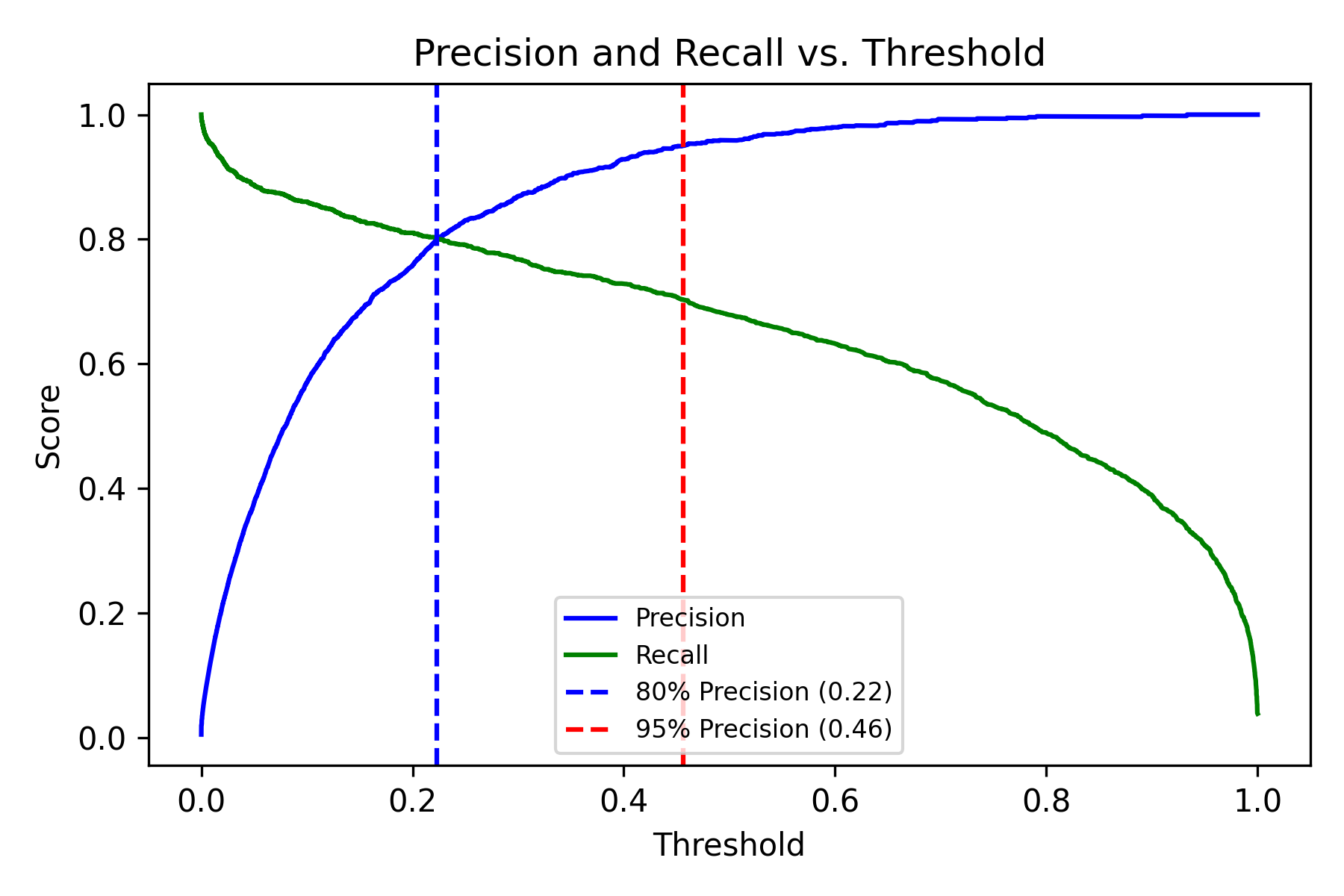}\hspace{5pt}%
    \includegraphics[width=0.5\linewidth]{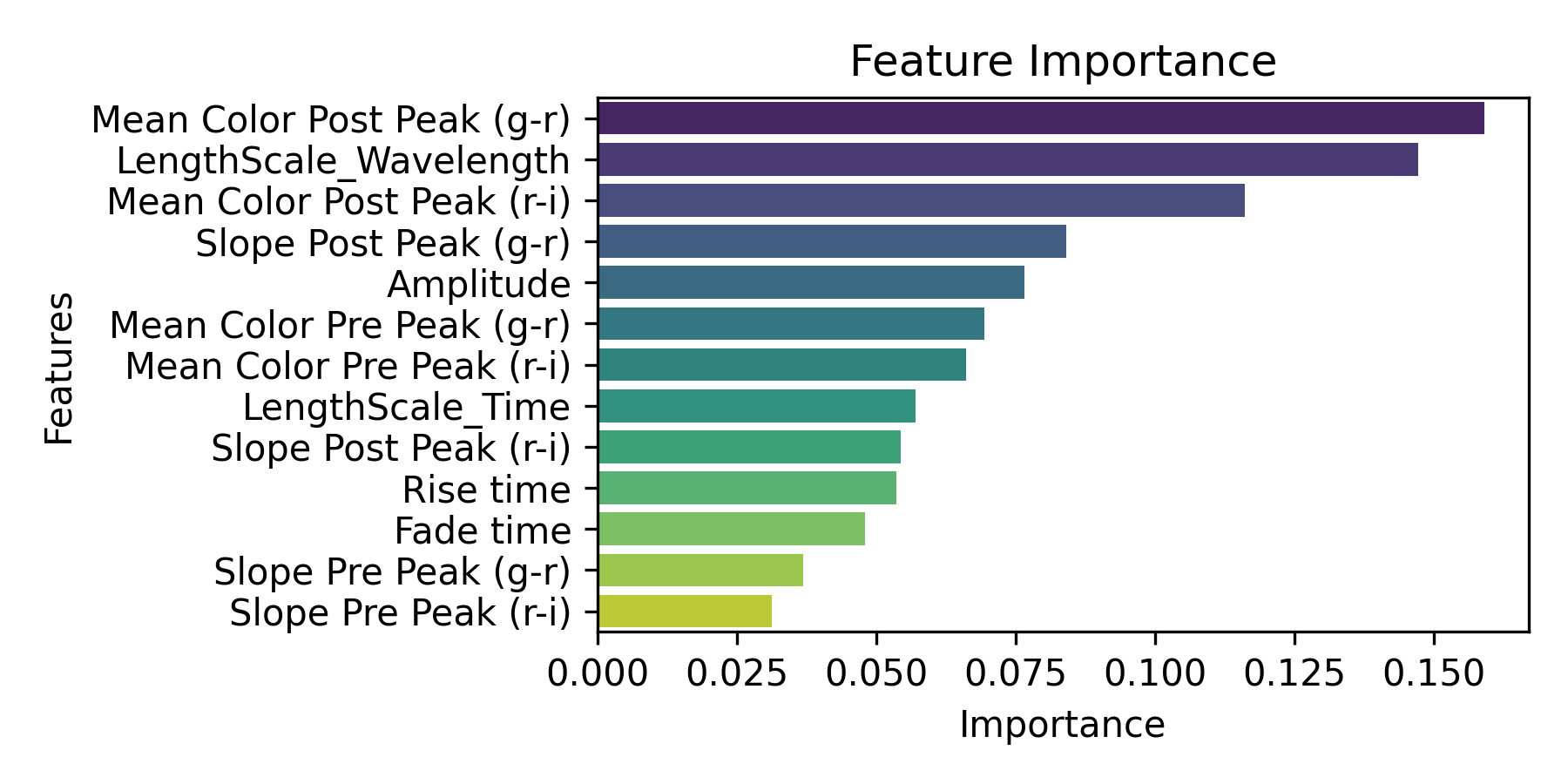}

    \vspace{10pt}

    \includegraphics[width=\linewidth]{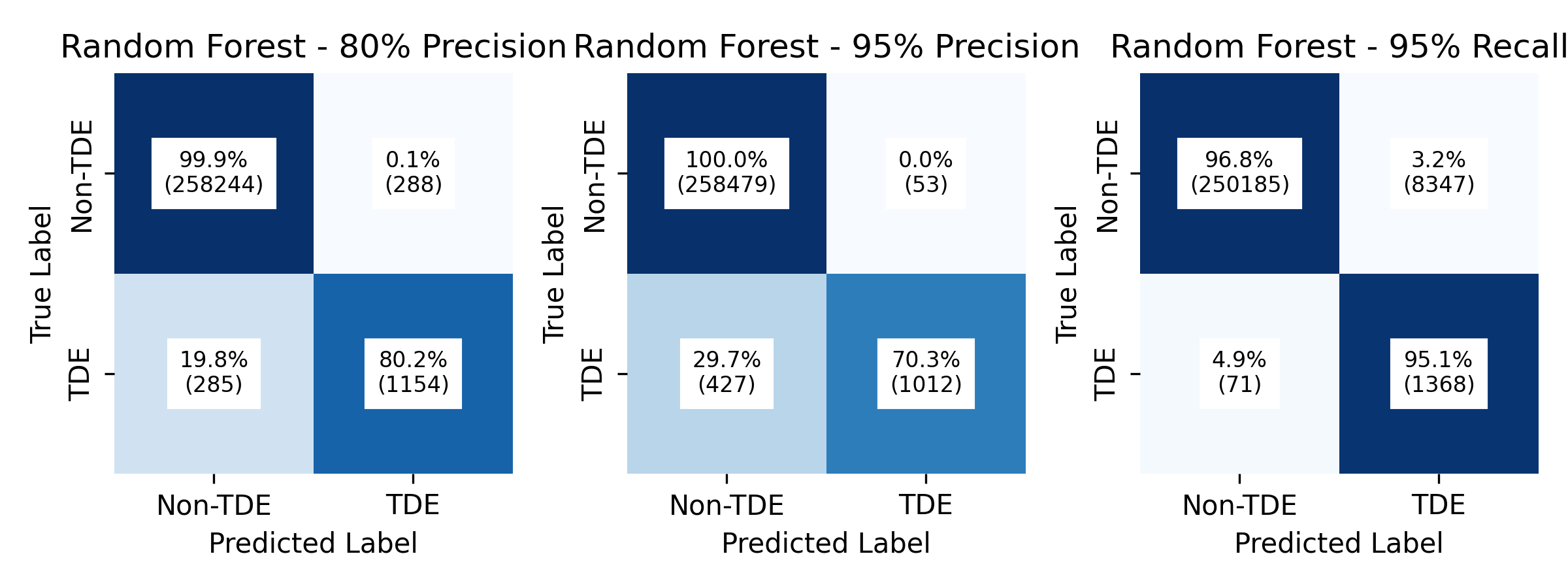}
    \caption{Random forest performance and outcomes. Upper left: Precision--recall curve.  Upper right: Feature importance.  Bottom: Confusion matrices at various thresholds.}
    \label{fig:RF_combined_fig}
\end{figure}

\begin{table}[htbp]
    \centering
    \caption{Optimized hyperparameters for the random forest classifier.}
    \label{tab:rf_params}
    \begin{tabular}{llp{0.55\linewidth}}
    \toprule
    \textbf{Parameter} & \textbf{Value} & \textbf{Description} \\
    \midrule
    \texttt{bootstrap} & False & Whether bootstrap samples are used when building trees. \\
    \texttt{class\_weight} & None & Adjust weights for classes (e.g., balanced). \\
    \texttt{max\_depth} & 48 & Maximum depth of each tree. \\
    \texttt{max\_features} & log2 & Number of features to consider at each split. \\
    \texttt{min\_samples\_leaf} & 3 & Minimum number of samples required at a leaf node. \\
    \texttt{min\_samples\_split} & 6 & Minimum number of samples required to split an internal node. \\
    \texttt{n\_estimators} & 918 & Number of trees in the forest. \\
    \bottomrule
    \end{tabular}
\end{table}

\end{appendix}

\end{document}